\documentclass[12pt]{article}

\usepackage{cite}

\usepackage{alltt}

\usepackage{graphicx}
\usepackage[figuresright]{rotating}

\def\Order#1{${\cal O}(#1$)}

\newcommand{\AmS}{{\protect\the\textfont2
  A\kern-.1667em\lower.5ex\hbox{M}\kern-.125emS}}

\begin{document}                     

\begin{titlepage}

\begin{flushright}
{\bf  CERN-TH/2002-194
}
\end{flushright}

\vspace{1mm}
\begin{center}
{\LARGE Precision predictions for $W$-pair production at LEP2$^{\dag}$}
\end{center}

\vspace{4mm}
\begin{center}
{\bf Wies\l{}aw P\l{}aczek}

\vspace{3mm}
{\em
Institute of Computer Science, Jagellonian University,\\
   ul. Nawojki 11, 30-072 Cracow, Poland,\\
{\em and}\\
CERN, TH Division, CH-1211 Geneva 23, Switzerland.\\
}
\end{center}

\vspace{14mm}
\begin{abstract}
Theoretical calculations for the $W$-pair production process at LEP2 
in terms of Monte Carlo event generators {\tt RacoonWW} and
{\tt KorlaW\&YFSWW3} are reviewed. The discussion concentrates on
precision predictions for the main LEP2 $WW$ observables. 
The theoretical precision of the above programs is estimated to be
$\sim 0.5\%$ for the total $WW$ cross section $\sigma_{WW}$, 
$\sim 5\,$MeV for the $W$-boson mass $M_W$,
and $\sim 0.005$ for the triple-gauge-boson coupling 
$\lambda=\lambda_{\gamma}=\lambda_Z$, which is
sufficient for the final LEP2 data analyses.
\end{abstract}

\vspace{15mm}

\begin{center}
{\it Talk given at the International Conference on High Energy Physics
     (ICHEP 2002),
     \\ Amsterdam, Netherlands, 24--31 July 2002; to appear in the
      proceedings.}
\end{center}

\vspace{17mm}
\footnoterule
\noindent
{\footnotesize
\begin{itemize}
\item[${\dag}$]
  Work partly supported by 
  the European Commission 5-th framework contract HPRN-CT-2000-00149.
\end{itemize}
}

\vspace{1mm}
\begin{flushleft}
{\bf CERN-TH/2002-194
\\  August 2002
}
\end{flushleft}

\end{titlepage}

\section{Introduction}

The process of $W$-pair production in electron--positron colliders is very
important for testing the Standard Model (SM) and searching for signals of
possible ``new physics''; see e.g.~Refs.~\cite{LEP2YR:1996,LEP2YR:2000}. 
One of the main goals when investigating this process at present and future
$e^+e^-$ experiments is to measure precisely 
the basic properties of the $W$ boson, 
such as its mass $M_W$ and width $\Gamma_W$. This process also 
allows for a study of the triple and quartic gauge-boson couplings at the tree
level, where small deviations from the subtle SM gauge cancellations can lead 
to significant effects on physical observables -- these can be signals
of ``new physics''.  On the other hand, the process $W$-pair production and 
decay constitutes a significant background to other processes, in particular 
to Higgs-boson production \cite{Mariotti:ICHEP02}. 

During the 1996--2000 LEP run, the four LEP2 experiments collected 
about 10,000 $W$-pair events each. This allows a test of 
the $W$-boson physics at the precision level of $\sim 1\%$ 
\cite{Grunewald:ICHEP02}. The main $WW$ observables investigated at LEP2
are (1) the total $WW$ cross section $\sigma_{WW}$, (2) a distribution of
the $W$ invariant mass, which is used to extract the $W$-boson mass $M_W$,
and (3) a distribution of the $W$ production polar angle $\cos\Theta_W$,
which is the most sensitive observable to the triple-gauge-boson couplings 
(TGCs). The expected experimental accuracies of the final LEP2 data analyses
for $\sigma_{WW}$, $M_W$ and the C- and P-conserving TGC 
$\lambda=\lambda_{\gamma}=\lambda_Z$ are given in Table~1.
In this table we also present requirements for the theoretical precision
of these quantities. For the theoretical errors ($\delta_{th}$) it is commonly
required that they should be smaller than the experimental ones 
($\delta_{ex}$) by a factor of at least $2$.\\[2mm]
Table~1:\\[-6mm]
\begin{center}
\begin{tabular}{||c|c|c||}
\hline\hline
Quantity & Final LEP2 $\delta_{ex}$ & Required $\delta_{th}$ \\
\hline
$\sigma_{WW}$ & $\sim 1\%$ & $\leq 0.5\%$ \\
$M_W$ & $\sim 30\,$MeV & $\leq 15\,$MeV \\
$\lambda$ & $\sim 0.01$ & $\leq 0.005$ \\
\hline\hline
\end{tabular}\\[2mm]
\end{center}
The final LEP2 data analyses need theoretical predictions in terms of
Monte Carlo event generator(s) (MCEG) that meet these precision
requirements.   

\section{Theoretical description of $W$-pair production}

The basic lowest-order process in which $W$ pairs were produced
at LEP2 is:
\begin{equation}
e^+ + e^- \rightarrow W^+ + W^- \rightarrow 
f_1 + \bar{f}_2 + f_3 + \bar{f}_4.
\label{WP:process}
\end{equation}
This can be described by three Feynman diagrams, called CC03:
two annihilation diagrams with $\gamma$ and $Z$ in the $s$-channel
(this is where the TGCs appear), and a conversion diagram with 
an electron neutrino in the $t$-channel (this contribution dominates
at LEP2 energies). And here the first theoretical problem appears.
Namely, these three diagrams alone are not gauge-invariant. In order
to obtain a gauge-invariant scattering amplitude, one has to include
a certain class of singly $W$-resonant diagrams. Therefore, the minimal
gauge-invariant subset of Feynman diagrams containing $W$ pairs is
the so-called CC11-class -- for more details see 
e.g.~\cite{WW-YR:1996}. This leads in practice to the necessity of
considering the full $4f$-process, i.e. $e^+ e^- \rightarrow 4f$,
which considerably complicates the theoretical description.

The next problem that appears in the theoretical description of this
process is the inclusion of a finite $W$-boson width $\Gamma_W$. This inclusion
is necessary to avoid singularities in the scattering amplitude coming
from $W$-boson propagators. Up to now there is no fully satisfactory
approach to deal with this in the general $4f$-process. Some simple
schemes, such as the ``fixed-width scheme'' and the ``running-width
scheme'' are known to violate gauge invariance. The schemes that
do not violate gauge invariance, such as the ``fermion-loop scheme'' 
or the ``complex-mass scheme'' possess other drawbacks, see 
e.g.~\cite{WW-YR:1996,4f-LEP2YR:2000}. In practice one commonly
uses the ``fixed-width scheme'', for which gauge-violating effects
are numerically negligible at LEP2 energies~\cite{WW-YR:1996}.

All the above problems increase dramatically when one goes to
\Order{\alpha} calculations. At the one-loop level, the number of Feynman
diagrams increases to several thousands per $4f$-channel (final state)
\cite{WW-YR:1996} and the inclusion of the finite $W$-boson width gets
much more difficult. 

One may ask the question: Do we need to go beyond the Born level?
So let us briefly discuss the various radiative corrections and their 
effects on the main LEP2 $WW$ measurements.
\\[2mm] \underline{\bf Pure QED correction:}\\
  QED corrections can be divided into a few classes: (1) initial-state
  radiation (ISR), i.e. photon radiation from incoming beams;
  (2) Coulomb correction -- electromagnetic interaction of slowly
  moving $W^+W^-$; (3) final-state radiation (FSR), i.e. photon radiation
  in $W$ decays; and (4) non-factorizable (NF) corrections corresponding to
  interconnections of various stages of the process. They affect the
  $WW$ observables in different ways. The most sizeable numerical effects
  come from the ISR. For $\sigma_{WW}$ they amount to between about $-20\%$
  near the $WW$ threshold and about $-5\%$ at $E_{CM} = 200\,$GeV
  (they depend stronlgy on $E_{CM}$). $M_W$ is affected at $\sim 10\,$MeV,
  but this may be enhanced by experimental reconstruction effects (kinematic
  fits, etc.). The TGC $\lambda$ is shifted by the ISR by $\sim 0.07$.
  The Coulomb correction modifies mainly $\sigma_{WW}$ at the level
  of $\sim 6\%$ near the $WW$ threshold. 
  The FSR influences considerably the $M_W$ measurement, at the
  level of $\sim 10$--$80\,$MeV, which strongly depends on experimental
  acceptances~\cite{yfsww3:1998b}. The NF corrections affect only $M_W$
  at the $\sim 1$--$5\,$MeV level, when treated 
  inclusively~\cite{Chapovsky:1999kv} (which is a good approximation 
  for LEP2 experimental acceptances). 
\\[2mm] \underline{\bf Electroweak (EW) corrections:}\\
  The leading EW corrections connected with an effective scale of
  the $W$-pair production and decay process can be taken into account
  by using the so-called $G_{\mu}$ scheme. This corresponds to parametrizing
  the cross section by the Fermi constant $G_{\mu}$ instead
  of the fine-structure constant $\alpha$. Numerically, this
  changes the overall normalization, i.e. $\sigma_{WW}$ by $\sim 15\%$. 
  Then the \Order{\alpha} non-leading (NL) EW corrections amount to
  $1$--$2\%$ for $\sigma_{WW}$ at LEP2 energies. They also affect the TGCs,
  e.g. $\lambda$ at the level of $\sim 0.01$--$0.02$. 
\\[2mm] \underline{\bf QCD corrections:}\\
  QCD corrections affect normalizations as well as events shapes of the
  hadronic $WW$ channels. They are usually accounted for by including
  the so-called naive QCD correction, i.e. the normalization factor 
  $\alpha_s(M_W^2)/\pi \simeq 3.8\%$ for each final-state quark pair
  and $2\alpha_s(M_W^2)/3\pi$ for $\Gamma_W$, in the parton-level
  calculation, while leaving the exclusive QCD effects to be modelled by
  dedicated MC packages, such as PYTHIA, etc. \\[-3mm] 

All the above effects are necessary in a MCEG for $W$-pair production at LEP2.
The most difficult are \Order{\alpha} NL EW corrections.
To date, there exist no complete one-loop calculations, 
even for the simplest CC11-type channels. 
Even if such calculations existed, they would probably be very complex
and slow in numerical computation. Therefore they would not be
useful for MC event generation. 
These are the reasons 
why some efficient approximations for the $WW$ process are necessary, 
at least for LEP2. 

In the case of $W$-pair production, one is interested in the doubly 
$W$-resonant process. 
Therefore one can make use of another expansion parameter,
which is $\Gamma_W/M_W \simeq 2.5\%$, and apply the so-called pole expansion,
i.e. an expansion about the complex pole corresponding to an unstable $W$.
This is the expansion in increasing powers of $\Gamma_W/M_W$, which
corresponds to decreasing powers of the resonance. Then, one usually
applies the so-called leading-pole approximation (LPA) in which only
the highest-pole (resonant) contributions are retained, i.e.\ terms
$\sim (\Gamma_W/M_W)^0$. For the $W$-pair production process this
means retaining only double-pole contributions, i.e. applying
a double-pole approximation (DPA). In this context LPA means
just DPA; however, LPA itself has a more general meaning.  
The resulting matrix element is gauge-invariant,
and the imaginary part of the pole position corresponds to the finite
$W$-boson width, see e.g.~\cite{stuart:1997,WW-YR:1996,4f-LEP2YR:2000}.
In general, the LPA is however not sufficient at the Born level,
because the error introduced by this approximation can be here
$\sim \Gamma_W/M_W$, i.e. $\sim 2.5\%$. It may also not be sufficient
for the leading corrections because of big-log enhancements, i.e. 
$\sim (\Gamma_W/M_W)(\alpha/\pi) \log(Q^2/m^2)$, where $Q$ is some
large momentum transfer and $m$ is a small fermion mass. 
Here lower-pole terms might be necessary for the desired experimental
precision. In practice, it is usually simpler to take the whole
$e^+e^- \rightarrow 4f$, including the leading corrections.
Then, for the genuine \Order{\alpha} NL EW corrections the uncertainty 
introduced by the LPA is $\sim (\Gamma_W/M_W)(\alpha/\pi) < 10^{-4}$,
therefore negligible for the LEP2 precision. 

The above solutions have been implemented in two MC event
generators: 
{\tt RacoonWW}~\cite{Denner:1999gp,Denner:1999kn,Denner:2000bj,Denner:2001vr}
and {\tt KoralW\&YFSWW3}~\cite{koralw:1995a,koralw:1995b,koralw:1997,%
koralw:1998,koralw:2001,yfsww2:1996,yfsww3:1998,yfsww3:1998b,yfsww3:2000a,%
yfsww3:2001}.
These programs are briefly described in the next two sections.

\section{The MC program {\tt RacoonWW}}

\noindent
The main features of the program {\tt RacoonWW} are:\\[-6mm]
\begin{itemize}
 \item Matrix elements for all $e^+e^- \rightarrow 4f$ and
       $e^+e^- \rightarrow 4f\gamma$ 
       processes in massless-fermion approximation. \\[-6mm]
 \item The ISR in the \Order{\alpha^2} leading-log (LL) approximation
       with soft-photon exponentiation through QED structure functions.\\[-6mm]
 \item Coulomb correction for off-shell $W$'s. \\[-6mm]
 \item The NF virtual corrections in the DPA.\\[-6mm]
 \item The virtual \Order{\alpha} NL EW corrections in the DPA
       (for which one-loop calculations of on-shell
        $WW$ production and decay are used).\\[-6mm]
 \item Two QCD-inspired methods of treating soft and collinear photon 
       singularities:
       dipole subtraction and phase-space slicing
       (to get a proper matching between virtual and real-photon corrections).
       \\[-6mm]
 \item Anomalous Triple and Quartic gauge-boson couplings.\\[-6mm]
 \item Multichannel MC algorithm for integration and event generation.\\[-6mm]
\end{itemize}
More details can be found in 
Refs.~\cite{Denner:1999gp,Denner:1999kn,Denner:2000bj,Denner:2001vr}. 

\section{The MC programs {\tt KoralW} and {\tt YFSWW3}}

{\tt KoralW} and {\tt YFSWW3} are actually two independent MCEGs,
with some specific and some common features, which are listed below.\\[3mm]
{\tt KoralW} {\bf specific features:} \\[-6mm]
\begin{itemize}
 \item The fully massive matrix element for all 
       $e^+e^- \rightarrow 4f$ processes (generated by the GRACE 
       System~\cite{GRACE2}).\\[-6mm]
 \item Two independent efficient multichannel MC algorithms
       for the full $4f$ phase space.\\[-4mm]
\end{itemize}
{\tt YFSWW3} {\bf specific features:}\\[-6mm]
\begin{itemize}
 \item Multiphoton radiation in the $WW$ production stage in the 
       Yennie--Frautschi--Suura (YFS) exponentiation scheme.\\[-6mm]
 \item The \Order{\alpha} NL EW corrections in the LPA
       (based on the \Order{\alpha} calculations for on-shell $WW$ production 
        of Refs.~\cite{fleischer:1989,kolodziej:1991}).\\[-4mm]
\end{itemize}
{\bf Common features:}\\[-6mm]
\begin{itemize}
 \item The ISR in the YFS exponentiation scheme up to 
       \Order{\alpha^3} LL with non-zero $p_T$ multi-photon radiation.\\[-6mm]
 \item Coulomb correction for off-shell $W$'s.\\[-6mm]
 \item Non-factorizable QED corrections in an inclusive approximation
       of the ``screened-Coulomb'' ansatz~\cite{Chapovsky:1999kv}.\\[-6mm]
 \item Anomalous TGCs (three parametrizations).\\[-6mm]
 \item The FSR at \Order{\alpha^2} LL generated by 
       {\tt PHOTOS}~\cite{photos:1994}
       (with non-zero $p_T$ photons). \\[-6mm]
 \item $\tau$ decays done by {\tt TAUOLA}~\cite{tauola:1993},
       quark fragmentation and hadronization managed by 
       {\tt JETSET}~\cite{jetset:1987}.\\[-6mm]
 \item The semi-analytical program {\tt KorWan} for the
       $WW$ process including leading corrections (for test, fits, etc.).
       \\[-5mm]
\end{itemize}

As can be seen from the above description, {\tt KoralW} is devoted
to the full $e^+e^- \rightarrow 4f$ process and it includes the
leading corrections, i.e. all except \Order{\alpha} NL. 
{\tt YFSWW3}, on the other hand, is dedicated to a precision description
of the signal $WW$ process, i.e. it includes all numerically important
radiative corrections to this process treated in the LPA, but does not
include the $4f$-background contributions. However, thanks to the common
features of the two programs, it is easy to correct the predictions of
one of them, so that the final results include all the necessary
contributions. In fact, the two programs implement special reweighting
routines for this purpose. Moreover, we managed to combine the results
of the two programs in a real-time execution (event-by-event) using
the Unix/Linux named (FIFO) pipe mechanism. This resulted in the
so-called Concurrent Monte Carlo (CMC) {\tt KoralW\&YFSWW3}.
This CMC works effectively as a single MCEG and provides the description
of the $W$-pair production process with all contributions/corrections
needed for the LEP2 precision -- for more details see Ref.~\cite{koralw:2001}.

\section{Theoretical precision of the main LEP2 $WW$ observables}

The precision requirements of the theoretical predictions for
some of the main LEP2 observables has been shown in Table~1.
Any MCEG to be used for the final LEP2 data analysis should satisfy
these requirements. Estimating the precision of theoretical calculations
is a difficult task, because it requires assessing the size of missing
higher-order contributions. In practice, this is done by comparing
different calculations (programs), switching on and off various 
contributions/corrections and looking at the effects for a given observable.
In some cases this can be supplemented with a simplified scale-parameter
analysis. All such investigations can give a hint on the size of
the missing effects in the theoretical description.

The programs {\tt RacoonWW} and {\tt YFSWW3} have been extensively 
cross-checked and compared for many observables,
see e.g.~\cite{4f-LEP2YR:2000}.
It has been found that for $\sigma_{WW}$ the two programs agree within
$0.3\%$ over the LEP2 energy range and they describe the LEP2 data 
very well~\cite{Malgeri:ICHEP02}.
From the comparisons between these two programs,
the comparisons with other calculations and investigations of various
effects, the theoretical precision for $\sigma_{WW}$ at LEP2 has been 
estimated at $\sim 0.5\%$~\cite{4f-LEP2YR:2000}.
Similar studies, although more involved, were also performed for
the $W$ mass $M_W$ and the TGC $\lambda$. In the case of the $W$ mass,
we performed $M_W$ fits using {\tt KorWan} to the invariant-mass 
distributions from {\tt KoralW\&YFSWW3} and {\tt RacoonWW},
where various effects have been switched on and off. We found,
for example, that results of the two programs differ in terms of
the fitted $M_W$ by $\le 3\,$MeV. We also performed a scale-parameter
estimation of the missing effects. From all these studies we 
estimated the theoretical precision of $M_W$ for the above programs
at $\sim 5\,$MeV. All this is described in detail in Ref.~\cite{Jadach:2001cz}.
A similar analysis for the TGC $\lambda$ was done in 
Ref.~\cite{Bruneliere:2002df}. In this case the fits of $\lambda$
to the $\cos\Theta_W$ distributions were performed using not
only {\tt KorWan} but also dedicated polynomial fitting functions.
These fitting functions were constructed using MC-genarated data
from {\tt KoralW\&YFSWW3} and {\tt RacoonWW}, not only at the parton level
but also including the ALEPH detector-simulation effects. 
Based on these studies, we
estimated the theoretical precision for $\lambda$ at $\sim 0.005$.
 
\section{Conclusions}
  There are two independent Monte Carlo programs for the precision predictions
   of $W$-pair production at LEP2: 
  {\tt RacoonWW}~\cite{Denner:1999gp,Denner:1999kn,Denner:2000bj,Denner:2001vr}
  and {\tt KoralW\&YFSWW3}~\cite{koralw:1995a,koralw:1995b,koralw:1997,%
  koralw:1998,koralw:2001,yfsww2:1996,yfsww3:1998,yfsww3:1998b,yfsww3:2000a,%
  yfsww3:2001}.
      They include the $4f$-background contributions as well as all the 
      necessary radiative corrections at the precision level required by 
      LEP2.
      The agreement between these programs for the main observables
      is within the required accuracy of the LEP2 experiments.
     From comparisons between these programs, comparisons with other
     calculations, and investigations of various effects, we have estimated
     the theoretical precision ($\delta_{th}$) 
     for three of the main LEP2 $WW$ observables: 
     $\delta_{th} \sigma_{WW}\sim 0.5\%$, $\delta_{th} M_W\sim 5\,$MeV
     and $\delta_{th}\lambda \sim 0.005$.
Comparing this with Table~1, it is clear that
this satisfies the LEP2 precision requirements. 
However, for the future linear colliders (LC) this is not sufficient
and further improvements in the theoretical calculations are necessary.



\end{document}